\documentclass[twocolumn,pra,superscriptaddress,nofootinbib,a4paper,longbibliography,floatfix]{revtex4-2}
\usepackage[utf8]{inputenc}
\usepackage{epsfig}
\usepackage{amsfonts}
\usepackage{amsmath}
\usepackage{amsthm}
\usepackage{amssymb}
\usepackage{amsthm}
\usepackage{dsfont}
\usepackage{bm}
\usepackage{enumerate}
\usepackage{mathtools}
\usepackage{color}
\usepackage{multirow}
\usepackage[normalem]{ulem}
\newcommand{\stkout}[1]{\ifmmode\text{\sout{\ensuremath{#1}}}\else\sout{#1}\fi}
\usepackage{latexsym}
\usepackage{mathrsfs}
\usepackage{natbib}
\usepackage{verbatim}
\usepackage[T1]{fontenc}
\usepackage{float}
\usepackage{graphicx}
\usepackage{xcolor}
\usepackage{physics}
\usepackage{soul}
\usepackage{subcaption}
\usepackage{float}
\usepackage[font=small,labelfont=bf,justification=justified]{caption}
\usepackage{xspace}
\usepackage{enumitem}
\usepackage{nicematrix}
\setlist[enumerate,1]{align=left}

\def \ecart {\noalign{\medskip}}

\renewcommand{\ket}[1]{|#1\rangle}
\renewcommand{\bra}[1]{\langle#1|}
\newcommand{\bracket}[3]{\langle#1|#2|#3\rangle}

\newcommand{\expect}[1]{\langle#1\rangle}

\makeatletter
\newcommand{\pushright}[1]{\ifmeasuring@#1\else\omit\hfill$\displaystyle#1$\fi\ignorespaces}
\newcommand{\pushleft}[1]{\ifmeasuring@#1\else\omit$\displaystyle#1$\hfill\fi\ignorespaces}
\makeatother

\usepackage{amssymb}
\usepackage{amsthm}
\usepackage{mathtools} 
\usepackage[customcolors]{hf-tikz} 
\usetikzlibrary{patterns}
\usetikzlibrary{matrix,decorations.pathreplacing}

\pgfkeys{tikz/mymatrixenv/.style={decoration={brace},every left delimiter/.style={xshift=8pt},every right delimiter/.style={xshift=-8pt}}}
\pgfkeys{tikz/mymatrix/.style={matrix of math nodes,nodes in empty cells,left delimiter={[},right delimiter={]},inner sep=1pt,outer sep=1.5pt,column sep=2pt,row sep=2pt,nodes={minimum width=20pt,minimum height=10pt,anchor=center,inner sep=0pt,outer sep=0pt}}}
\pgfkeys{tikz/mymatrixbrace/.style={decorate,thick}}

\usepackage[colorlinks=true,linkcolor=magenta,citecolor=magenta,urlcolor=blue]{hyperref}

\begin{document}
	

\title{Detecting high-dimensional entanglement with simple measurements}

\author{Suraj Goel}
\email{s.goel@hw.ac.uk}
\affiliation{Institute of Photonics and Quantum Sciences, Heriot-Watt University, Edinburgh, UK}

\author{Alexander Bernal}
\affiliation{Physics Department and NanoLund, Lund University, Box 118, 22100 Lund, Sweden.}

\author{Gabriele Cobucci}
\affiliation{Physics Department and NanoLund, Lund University, Box 118, 22100 Lund, Sweden.}

\author{Will McCutcheon}	
\affiliation{Institute of Photonics and Quantum Sciences, Heriot-Watt University, Edinburgh, UK}

\author{Mehul Malik}	
\affiliation{Institute of Photonics and Quantum Sciences, Heriot-Watt University, Edinburgh, UK}

\author{Armin Tavakoli}
\email{armin.tavakoli@fysik.lu.se}
\affiliation{Physics Department and NanoLund, Lund University, Box 118, 22100 Lund, Sweden.}

\begin{abstract}
The standard benchmark for high-dimensional entanglement is  the number of dimensions in which entanglement must be present in order to generate the state. This is called the Schmidt number and its detection is usually based on implementing an appropriate set of local basis measurements. However, as quantum technology brings increasingly large physical dimensions within reach, the implementation of such measurements typically becomes more costly. Here, we develop a scheme for detecting Schmidt numbers based only on sequences of  single-qubit observables. These measurements are simpler to implement as they require only low-depth quantum circuits. Using up to sixteen-dimensional photonic spatial mode entanglement and multi-plane light conversion technology, we demonstrate how it simplifies setup complexity and successfully detects the maximal (or close-to-maximal) Schmidt number. Our results reveal that simple and more scalable measurements are sufficient to detect  high-dimensional entanglement properties.
\end{abstract}

\date{\today}

\maketitle

\textit{Introduction.---} High-dimensional systems are increasingly interesting for quantum information science, ranging from applications in high-capacity and noise-robust quantum communication to quantum computing protocols with reduced complexity~\cite{malik2026high}. However, establishing their entanglement properties is a central challenge. A core question, both from a theoretical and experimental perspective, is how to certify the entanglement dimensionality in practically convenient ways. 

\begin{figure}[!ht]
	\centering
	\includegraphics[width=1\columnwidth]{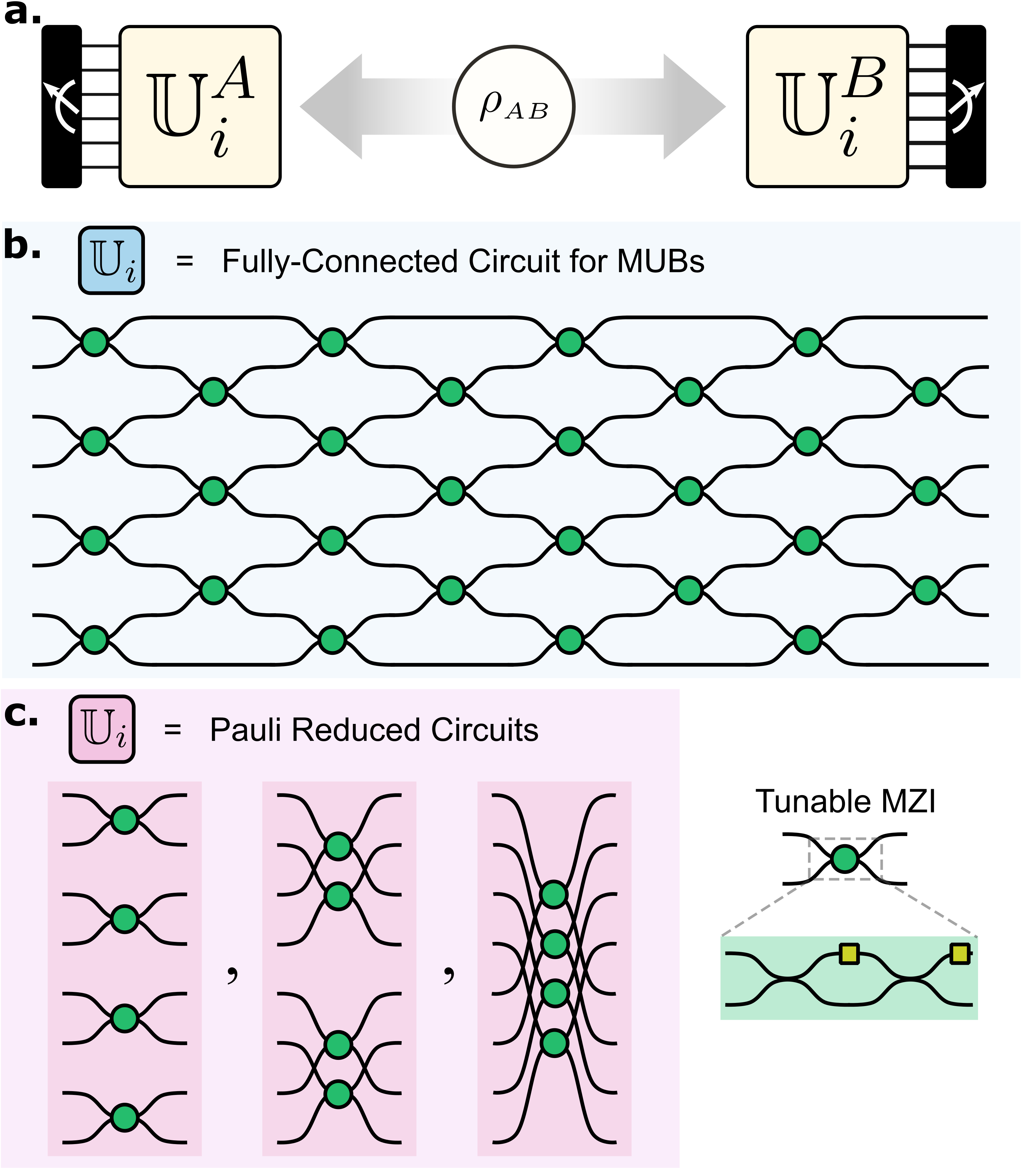}
	\caption{\textbf{Simpler measurements for entanglement witnessing.}
    (a) When a high-dimensional quantum state is distributed between two parties, they perform joint measurements to verify the entanglement dimensionality. (b) Schematic of conventional measurements in mutually unbiases bases~(MUBs) which correspond to a fully-connected optical circuit. (c) 
    Our method requires lower-depth circuits connected in a few different pairing configurations. Mach-Zehnder interferometer~(MZI) based optical circuits are shown used to measure path-encoded quantum states.}\label{fig1}
\end{figure}

The entanglement dimensionality --- formally called the Schmidt number \cite{Terhal2000} --- is computationally hard to evaluate \cite{Gharibian2008}. Therefore, focus has been on finding sufficient criteria for the Schmidt number. These rely on either the entire density matrix \cite{Weilenmann2020, Morelli2024, Liu2024}, randomised measurements \cite{Wyderka2023, Liu2023} or well-tailored witness tests \cite{Wyderka2023b, Cobucci2024, Morelli2023, Bavaresco2018}. Figure~\ref{fig1}a depicts the experimental scenario for implementing all three criteria, where local unitary transformations $U_i$ followed by a multi-outcome measurement are used at each party. 
The first approach is suitable only for low-dimensional systems, as it needs state tomography to be performed in order to obtain the full density matrix. The second requires the capacity to implement arbitrary unitary transformations on the $d$-dimensional local systems. The third, however, just needs the experimenter to make a few well-chosen measurements to deduce the Schmidt number. A common class of witnesses are  based on measuring fidelity with the maximally entangled state \cite{Terhal2000}, which is frequently used in experiments; see e.g.~\cite{Bavaresco2018, Fickler2014, Malik2016, Herrera2020, Hu2020, Hu2025, Moreno2026}.

Schmidt number witnesses commonly use local measurements that involve transformations between mutually unbiased bases (MUBs) or variations thereof \cite{Bavaresco2018, Morelli2023}.  
However, the complexity of these transformations scales poorly with its dimension. 
This complexity can be quantified by the depth of the optical circuit of the corresponding transformation. 
For conventional path-encoded Mach-Zehnder Interferometer~(MZI) based circuit designs, this depth is defined by the the number of interferometers a photon encounters while traversing the longest path in the circuit~\cite{clements_optimal_2016}.
For example, MUB transformations on a $d$-dimensional path-encoded state require a photonic circuit with a depth of $d$ (see Fig.~\ref{fig1}b for an example circuit with $d=8$).
A large circuit depth equates to a large number of circuit elements such as phase-shifters and beam-splitters. As the circuit depth increases, imperfections in these elements and the associated control complexity have a detrimental impact on the quality of the implemented transformations~\cite{burgwal2017using}. While reconfigurable circuits in up to $d=24$ have been demonstrated recently~\cite{barzaghi202524}, scaling to larger dimensions remains a significant challenge. Thus, there is a strong practical motivation for finding low-depth implementations of high-dimensional  entanglement detection. 

Here, we theoretically develop and experimentally demonstrate a Schmidt number witness method that uses only a small number of measurements, where each measurement can be factored into multiple parallel measurements of lower dimension. 
This enables each  high-dimensional observable to be implemented as a function of multiple two-dimensional Pauli-transformations, requiring  significantly lower circuit depth. 
We use these circuits to witness high-dimensional entanglement: Fig.~\ref{fig1}c illustrates a set of three reduced, unit-depth circuits sufficient for certification, in contrast to the fully connected $d=8$ 
path-encoded circuit required for conventional MUB-based certification (Fig.~\ref{fig1}b).
Our witnesses can both detect unfaithful entanglement and be used for fidelity estimation with the maximally entangled state.

We showcase the utility of our technique by generating high-dimensional photonic entanglement in the transverse-spatial degree-of-freedom and witnessing it using measurements of significantly reduced complexity. This platform is particularly well-suited to our method: the complexity of high-dimensional spatial-mode transformations degrades their quality~\cite{goel2025quantum}, which in turn limits the entanglement dimensionality that can be certified, while the fact that spatial modes can be rearranged into multiple configurations makes constructing multiple two-dimensional bases straightforward.
Using only unit-depth circuits, we certify the maximal entanglement dimension in an 8×8-dimensional Hilbert space, whereas generic fully connected operations on the same dimension would require a circuit depth of eight.
In addition, we certify $13$-dimensional entanglement in a 16×16-dimensional Hilbert space using circuits of depth $1.5$.
For our implementation, these $1.5$ depth circuits are in principle deterministic for certifying systems of arbitrary dimension. We further discuss extensions to higher dimensions and other photonic degrees of freedom. Our work opens the door to scalable  high-quality certification of high-dimensional entanglement with significantly reduced experimental complexity.

\textit{Preliminaries.---} Let $\psi_{AB}=\ketbra{\psi}$ be a pure bipartite state of local dimension $d$. Its Schmidt number is the rank of its marginals, i.e.~$k(\psi)=\rank\left(\psi_{A}\right)$. The standard way to extend this definition to mixed states, $\rho_{AB}$, is to consider pure-state decompositions, $\rho_{AB}=\sum_i q_i \ketbra{\psi_i}$, where $\{q_i\}_i$ is a probability distribution, and consider the highest Schmidt number appearing among the pure states in the decomposition, namely  $\max_i k(\psi_i)$. This quantity must then be minimised over all possible decompositions of $\rho_{AB}$. Thus, the Schmidt number is \cite{Terhal2000},
\begin{align}\nonumber
k(\rho_{AB})\equiv \min_{\{q_i\},\{\psi_i\}} \Big\{&k_\text{max}: \quad \rho_{AB}=\sum_i q_i \psi_i\\
&\text{and}\quad k_\text{max}=\max_i k(\psi_i)\Big\}.
\end{align}
Computing $k(\rho_{AB})$ is hard but a simple lower bound can be obtained from the fidelity of $\rho_{AB}$ with a maximally entangled state, namely $F(\rho_{AB})= \bracket{\phi^d}{\rho_{AB}}{\phi^d}$, where $\ket{\phi^d}=\frac{1}{\sqrt{d}}\sum_{i=0}^{d-1}\ket{ii}$. Any state with Schmidt number at most $k$ satisfies  \cite{Terhal2000}
\begin{equation}\label{fidcriterion}
    F(\rho_{AB})\leq\frac{k}{d}.
\end{equation}
Thus, if this inequality is violated, the Schmidt number is at least $k+1$.	The advantage of this method is that it applies to any pair $(k,d)$. Its main drawbacks are that  many high-dimensional entangled states cannot be detected through the fidelity criterion \cite{Weilenmann2020}, and that measuring the fidelity corresponds to $d+1$ product basis measurements. With fewer measurements, one can still infer lower  bounds on $F(\rho_{AB})$ \cite{Bavaresco2018, Morelli2023}. 

\textit{Witness method.---} We propose a Schmidt number witness method that requires only a modest number of observables, all of which can be implemented as tensor products of standard single-qubit Pauli observables. In other words, each high-dimensional measurement effectively reduces to strings of qubit building-blocks. Despite this simple structure, the method can detect the Schmidt number of states that are insensitive to the fidelity criterion \eqref{fidcriterion}, including bound entanglement. Moreover, it can also provide lower bounds on the fidelity $F(\rho_{AB})$ which is often a relevant benchmark of an entanglement source.

Let us begin with discussing the type of measurements used in the witness. To define them let $(X,Y,Z)$ be the standard Pauli observables and set $S_{1,1}=Z$, $S_{1,2}=X$ and $S_{1,3}=Y$. From this, we can define  strings of Pauli observables  by the recursion rule
\begin{align}\nonumber \label{eq:PS_defn}
& S_{n+1,j}=S_{n,j}\otimes Z, \qquad \text{for } j=1,\ldots,2n+1\\\nonumber
& S_{n+1,2n+2}=\openone^{\otimes n} \otimes X, \\
&  S_{n+1,2n+3}=\openone^{\otimes n} \otimes Y.
\end{align}
Thus, there are $2n+1$ strings, each comprised of $n$  Pauli observables. We denote by $G_k^{n,j}$ the $k$'th Pauli observable in the $j$'th string of length $n$. The key property of these Pauli-strings is that they pairwise anti-commute, i.e.~$\{S_{n,i},S_{n,j}\}=0,
$ $\forall i\neq j$, as shown in Appendix~\ref{AppA}. For a given $n$, no larger set of anti-commuting Pauli-strings is possible than the one given above \cite{Sarkar2019}.
	
 \begin{figure}
 	\centering
 	\includegraphics[width=1\columnwidth]{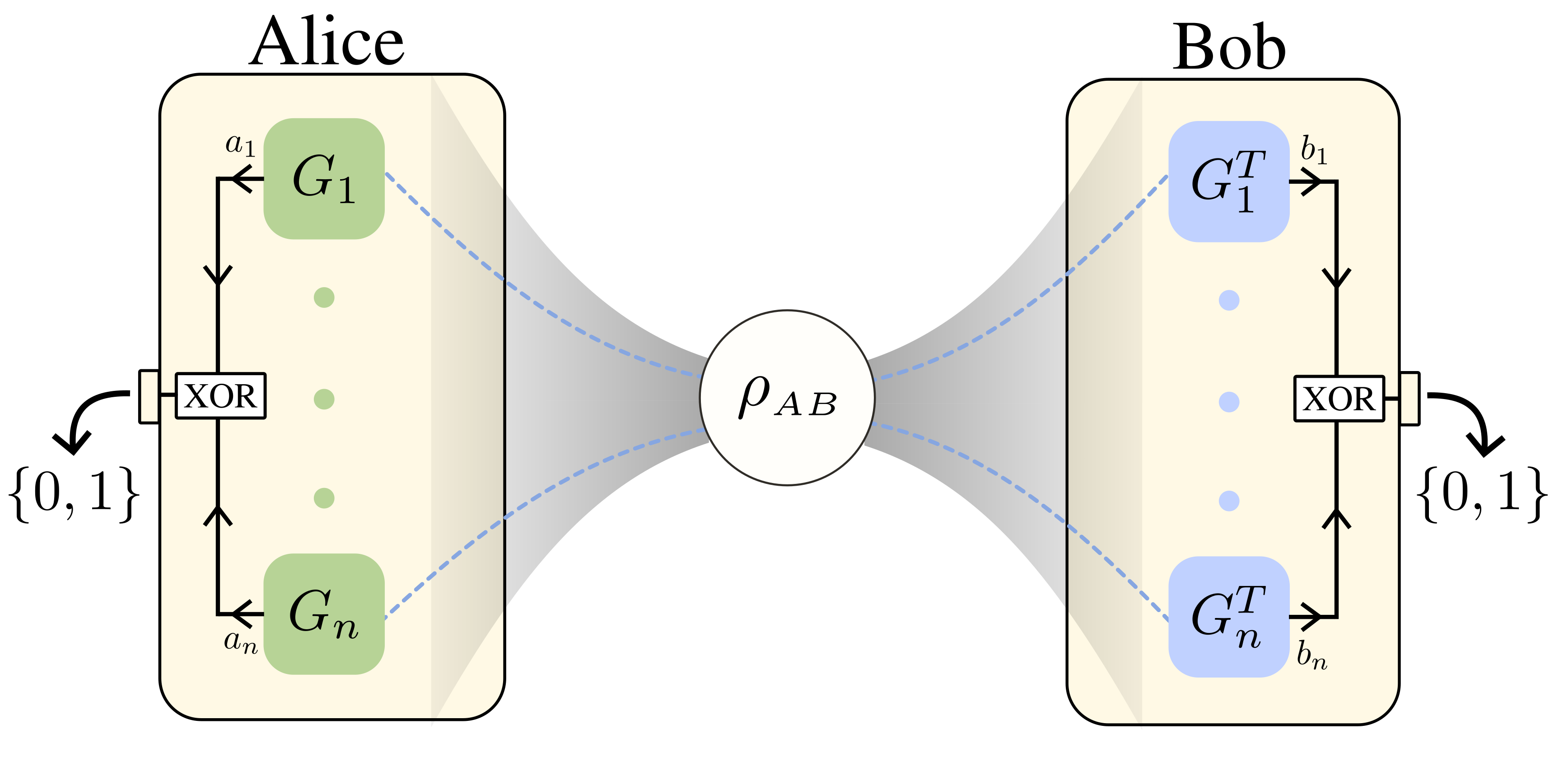}
 	\caption{Alice and Bob each receive a share of $\rho_{AB}$  consisting of $n$ qubits. They measure each qubit using an appropriate Pauli observable, $G$ ($G^T$). From the respective string of outcomes, $\lbrace a_{1},\dots,a_{n} \rbrace$ and $\lbrace b_{1},\dots,b_{n} \rbrace$, they each compute the parity and return it as their respective final output.}\label{fig_scheme}
 \end{figure}

We now use these observables for witnessing the Schmidt number of a bipartite quantum state, $\rho_{AB}$, whose local dimension is $d=2^n$.  Alice and Bob each view their share of the state as a collection of $n$ qubits. By applying the Pauli-string observables to their respective shares, they can effectively measure each qubit independently and classically wire the outcomes of each single-qubit measurement to compute the final binary outcome of the $n$-qubit measurement; see Fig~\ref{fig_scheme}. Specifically, Alice and Bob both select an index $j\in\{1,\ldots,2n+1\}$. Alice measures $S_{n,j}$ and Bob measures the transposed string, $(S_{n,j})^T=\bigotimes_{k=1}^n (G^{n,j}_k)^T$. The figure of merit is the joint expectation value of their observables, summed over all possible choices of $j$. Hence, the witness parameter is 
\begin{equation}\label{witness}
W_n(\rho_{AB}) \equiv \sum_{j=1}^{2n+1} \expect{S_{n,j} \otimes (S_{n,j})^T}_{\rho_{AB}}.
\end{equation}	
The maximal value of $W_n$ is achieved with the maximally entangled $d$-dimensional state. To see that, decompose the state as  $n$ copies of the maximally entangled qubit state, namely $\phi^d_{AB}= \phi^+_{A_1B_1}\otimes\ldots \otimes \phi^+_{A_nB_n}$, where $\phi^+=\ketbra{\phi^+}$ and  let $\ket{\phi^+}=\frac{1}{\sqrt{2}}\left(\ket{00}+\ket{11}\right)$ and $A_1,\ldots,A_n$ ($B_1,\ldots,B_n$) denote the qubit subsystems of Alice (Bob). Then,  $W_n(\phi^d)=\sum_j \prod_{k} \expect{G^{n,j}_k \otimes (G^{n,j}_k)^T}_{\phi^+}=2n+1$.  Here, we  used that $\expect{O \otimes O^T}_{\phi^+}=1$ for every linear operator, $O$, satisfying  $O^2=\mathds{1}$.  

In order to witness the Schmidt number of $\rho_{AB}$, we must determine bounds on the largest value of $W_n$ achievable by states whose Schmidt number is at most $k$. To this end, we begin with estimating the fidelity. In Appendix~\ref{AppB_2}, we use an induction argument to prove that  $F(\rho_{AB})\geq \frac{1}{4}(W_n+3-2n)$. Inserting this fidelity bound into  Eq~\eqref{fidcriterion}, we obtain the Schmidt number witness
\begin{equation}\label{fidestimate}
    W_n(\rho_{AB}) \leq 2n-3+\frac{k}{2^{n-2}}.
\end{equation}
While this is the optimal bound for fidelity estimation via $W_n$, some values of $W_n$ that satisfy the above inequality are still incompatible with states with Schmidt number at most $k$. This is because $W_n$ can detect the Schmidt number of states that are out of the reach of fidelity-based methods. For instance, in Appendix~\ref{AppB_1} we prove that all states with Schmidt number $k$ satisfy 
\begin{equation}\label{alexbound}
    W_n(\rho_{AB})\leq 2k-1.
\end{equation}
When $k<n$, this is a better bound than our fidelity-based criterion in Eq~\eqref{fidestimate}.  Bound entanglement can violate this inequlity, and for $k=1$ it is closely related to the uncertainty relations derived in \cite{Toth2005}.

\setlength{\tabcolsep}{4pt}
\begin{table}[t!]
\resizebox{\columnwidth}{!}{%
	\begin{tabular}{c||c|c|c|c|c|c|c|c}
		$k$  & 1 & 2 & 3 & 4 & 5 & 6 & 7 & 8 \\ \hline\hline
		$n=2$ 
		& \begin{tabular}[c]{@{}c@{}}$1$ \\  $1$\end{tabular}
		& \begin{tabular}[c]{@{}c@{}}$3$ \\  $3$\end{tabular}
		& \begin{tabular}[c]{@{}c@{}}$1+2\sqrt{2}$ \\ $4$\end{tabular}
		& \begin{tabular}[c]{@{}c@{}}$5$ \\ $5$\end{tabular}
		& -
		& -
		& -
		& - \\ \hline
		$n=3$ 
		& \begin{tabular}[c]{@{}c@{}}$1$ \\ $1$\end{tabular}
		& \begin{tabular}[c]{@{}c@{}}$3$ \\ $3$\end{tabular}
		& \begin{tabular}[c]{@{}c@{}}$1+2\sqrt{2}$ \\ $9/2$\end{tabular}
		& \begin{tabular}[c]{@{}c@{}}$5$ \\ $5$\end{tabular}
		& \begin{tabular}[c]{@{}c@{}}$1 + \sqrt{10+2\sqrt{17}}$ \\ $11/2$\end{tabular}
		& \begin{tabular}[c]{@{}c@{}}$3+2\sqrt{2}$ \\ $6$\end{tabular}
		& \begin{tabular}[c]{@{}c@{}}$1+2\sqrt{7}$ \\ $13/2$\end{tabular}
		& \begin{tabular}[c]{@{}c@{}}$7$ \\ $7$\end{tabular}
	\end{tabular}
    }
	\caption{Numerically optimised value of $W_n$ (upper) and the best upper bound selected from Eqs~\eqref{fidestimate} and \eqref{alexbound} (lower), for $n=2,3$ and every $k$. The value conjectured optimal via Eq~\eqref{Subset_Opt} coincides with the numerical results.}
	\label{Tab1}
\end{table}

\begin{figure*}[th]
\centering
\includegraphics[width=1.9\columnwidth]{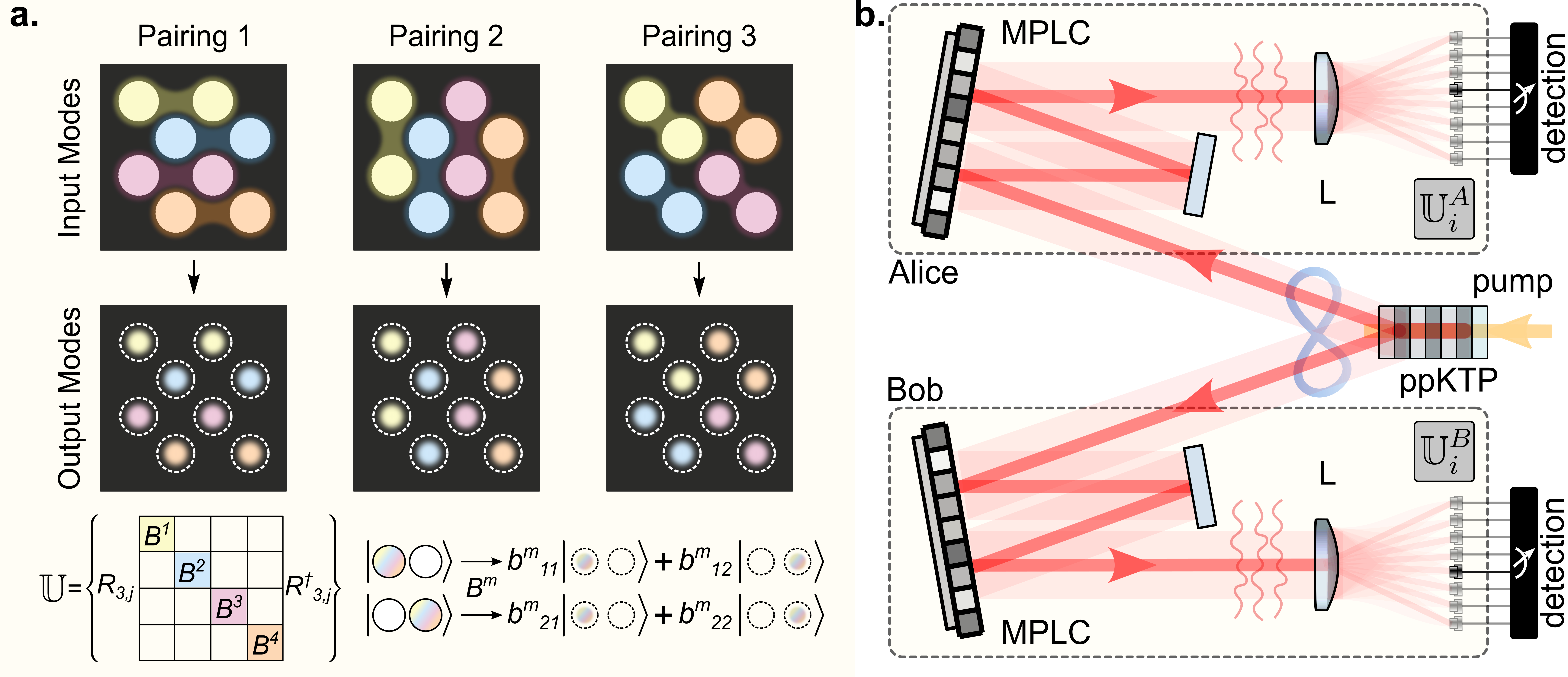}
\caption{\textbf{Illustration of the experimental scheme.} (a)~Measurement circuit transforms macro-pixel modes to spatially separated Gaussian spots. The eight input pixels~(and output spots) are arranged in a configuration that allows for nearest neighbour operations in three different pairing configurations ($R_{3,j}$), effectively needing four parallel two-mode operations. The measurement operation~$\mathbb{U}$ corresponding to the Pauli string observables for $d=8$ can be expressed as a block diagonal matrix, with blocks $B^{j}$, each operating on a pair of macro-pixels resulting in a pair of outcomes. (b)~Schematic illustration of the experimental setup, where the entangled pair of photons are generated using spontaneous parametric downconversion~(SPDC), and measured by independent parties (Alice and Bob), each using a two-plane multi-plane light converter~(MPLC) performing the relevant operations. 
} \label{fig_setup}
\end{figure*}

However, even stronger bounds are possible. Following the procedure outlined in Appendix~\ref{AppC}, we have numerically optimised the witness for $n=2,3$ for every choice of $k$. The results are shown in Table~\ref{Tab1}: they are typically lower than our analytical bounds, but the gap tends to narrow for larger Schmidt numbers. Interestingly, the results are always found to correspond to pure states whose Schmidt basis is the computational basis. Based on this observation, we let $T_k$ denote the set of all such pure states with Schmidt rank $k$. We prove in Appendix~\ref{AppD} that  
\begin{equation}\label{Subset_Opt}
\begin{aligned}
&\max_{\psi_{AB}\in T_k} W_n(\psi_{AB})=1+2\mu(n,k), \quad \text{where}\\ \ecart
    &\mu(n,k)=\max_{|S|= k} \lambda_{\max}(\text{Adj}(Q_n[S])).
\end{aligned}
\end{equation}
Here, $Q_n$ is the $n$-dimensional hypercube graph, which has $2^n$ vertices, and the maximisation is over any subgraph $S$ with $k$ vertices. The maximisation concerns the largest eigenvalue of the adjacency matrix of the subgraph. For general graphs, such maximisations are NP-hard problems \cite{Karp1972,MAGDONISMAIL201735}, and for the specific case of the hypercube graph no closed algebraic formula for $\mu(n,k)$ is known \cite{BOLLOBAS2018125}. Nevertheless, once an exhaustive exploration of all $k$-size subgraphs $S$ is performed, the value $\mu(n,k)$ is exactly computable and methods have been developed to speed up this computation \cite{BOLLOBAS2018125}. Using this method, we have evaluated  $1+2\mu(n,k)$ up to $n=5$ for every $k$. These match the  aforementioned numerical results for $n=2,3$, and the values for larger $n$ are displayed in Table~\ref{Tab2} in Appendix~\ref{AppD}. We conjecture that Eq~\eqref{Subset_Opt} is the optimal bound for the Schmidt number witness.  Notably, independently of the veracity of the conjecture, we can use these results to prove that the bounds \eqref{alexbound} and \eqref{fidestimate} are optimal for $k=2$ and $k=2^{n-1}$ respectively. The latter is particularly relevant, since it corresponds to certifying that all qubit pairs in Alice and Bob's larger state are entangled. To see this,  we can select the specific subgraph $S=Q_m$ associated with $k=2^m$. This gives the lower bound $\mu(n,k)\geq \lambda_{\max}(\text{Adj}(Q_n[Q_m]))= m$ which for $m=1, n-1$ matches Eq~\eqref{alexbound} and Eq~\eqref{fidestimate} respectively.

\textit{Results.---} To demonstrate the simplicity of this method, we use it to experimentally certify the dimensionality of a biphoton state entangled in the transverse-spatial degree-of-freedom. To efficiently characterise a high-dimensional state, one needs generalised multi-outcome measurements. Here, each Pauli string observable $S_{n,j}$ can be measured by recasting its eigenvectors as columns of a block-diagonal unitary matrix $U_{n,j}$ followed by a modal permutation operation $R_{n,j}$. 
In general, for $(n,d=2^n)$, one needs at most $n$ different permutations to realise all required Pauli-string measurements~(See Appendix~\ref{App_block_diagonal}). To perform these measurements on our bi-partite state, we use a pair of multi-plane light converters~(MPLC) as the optical circuits. 
For optical circuits such as MPLCs, which are designed in a top-down scheme, the circuit depth corresponds to half of the number of phase-planes used~\cite{goel_inverse_2024,clements_optimal_2016}.

We generate the biphoton state in the telecom band~($1550$~nm) by the means of spontaneous parametric downconversion~(SPDC) by pumping a periodically poled potassium titanyl phosphate~(ppKTP) crystal with a $775$~nm laser source, as illustrated in Fig.~\ref{fig_setup}b. 
The state is then distributed between two parties~(See Appendix~\ref{App_Exp} for experimental details), namely Alice and Bob, who employ an MPLC each as their optical circuit to simultaneously perform the operations corresponding to a chosen observable, as well as sort the measurement outcomes into the bespoke output basis.

We first fix $n=3$ and focus on the case of an entangled state with local dimension $d=8$. To use our witness and perform the $2n+1=7$ Pauli string observables, we need 3 modal permutations $R_{3,j}$. We choose a geometry of $8$ macro-pixel spatial modes where nearest-neighbour pairings are possible for each pixel in three different ways (Fig.~\ref{fig_setup}a) and associate each modal permutation with a certain pairing configuration. We use these macro-pixel modes as the input computational basis for the measurement circuit. 
Correspondingly, the outcomes are spatially separated Gaussian modes distributed in accordance to the same geometry. 
This geometry of modes enables us to implement all the required Pauli-string observables via MPLCs designed using only two phase planes: this corresponds to a circuit depth of one. 
In a similar way, we also implement the observables needed for the case of an entangled state of local dimension $d=4$ (which corresponds to $n=2$). This requires exactly two pairing configurations~(See Appendix~\ref{App_data} for more details).

For $d=4$, we measure a witness value of $W = 4.92 \pm 0.05$, which certifies a fidelity with the maximally entangled state of at least $F=98.12\pm 1.27\%$. This corresponds to the maximal Schmidt number, $k=4$. 
Next, for $d=8$, we measure a witness value $W = 6.52 \pm 0.07$. This certifies a fidelity with the maximally entangled state of at least $F=88.10\pm 1.67\%$.
By itself, the statistical errors would limit the Schmidt number to be $k=7$. 
However, as we know that the detection power of the method can go beyond that of fidelity estimation, we observe that the improved bound for $(n,k)=(3,7)$ in Table~\ref{Tab1} is violated, thereby certifying a maximal Schmidt number of $k=8$. 
The errors in the witness and fidelity are measured to three standard deviations from statistics obtained by performing Monte-Carlo error propagation on the data assuming Poissonian statistics of the measured correlations.
The measured coincidence matrices corresponding to each Pauli string observables for $n=\{2,3\}$ are shown in Appendix~\ref{App_data}.

\begin{figure}[!ht]
\centering
\includegraphics[width=\columnwidth]{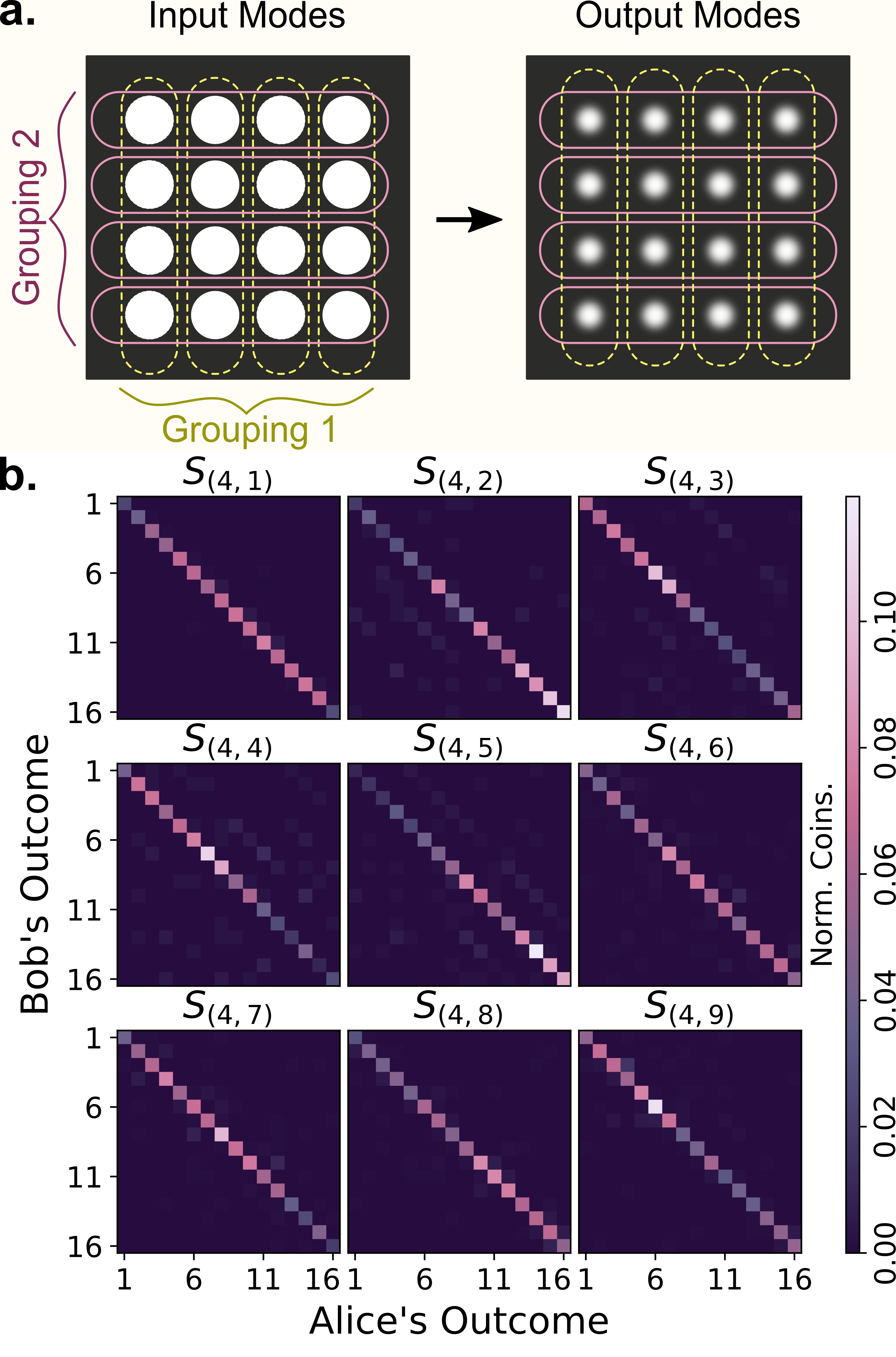}
\caption{\textbf{Entanglement detection in a 16-dimensional subspace} (a)~16 macro-pixel input modes are chosen in a 4-column grid to be transformed into a grid of spatially separated Gaussian spots such that two configurations of 4-mode groups are possible in enabling nearest neighbour interaction. The nine Pauli string operators can be expressed as four parallel $4\times4$ operations in the bespoke configurations. These operations are implemented using a three-plane MPLC. (b)~Measured normalised coincidence (Norm. Coins.) matrices between Alice's and Bob's outcomes corresponding to each Pauli-string observable for $d=16$ dimensional state. The measured data certifies the bound for $13$-dimensional entanglement. }\label{fig_16dim}
\end{figure}

Next, we construct local measurements for $n=4$, corresponding to local dimension $d=16$ of the entangled state. 
This case requires four pairing configurations of the optical modes, which is a challenging to construct such that there is nearest neighbour pairings between spatial modes arranged in a 2D-transverse spatial plane. 
Instead, we factorise these measurements with four 
parallel $4$-mode transformations, which can now be done with only two configurations. 
This method of grouping modes has been explored previously for reducing the complexity of implementing high-dimensional measurements for QKD~\cite{Lib25}.
We employ an MPLC with three phase-planes to implement the measurement operations to accommodate for parallel $4$-dimensional transformations. 
Fig.~\ref{fig_16dim}a shows the geometry of the $16$ macro-pixel spatial modes with two different groups of four neighbouring pixel modes.
We program the measurement circuits  corresponding to $2n+1=9$ Pauli-string observables with the bespoke macro-pixel modes as inputs, and relevant spatially separated Gaussian modes as outputs.

For $d=16$, we measure a witness value of $W = 8.12 \pm 0.11$, which certifies a fidelity with the maximally entangled state of at least $F=78.12\pm 2.69\%$.
This certifies a Schmidt number $k=13$.
The measured coincidence matrices for each string for $d=16$ is shown in  Fig.~\ref{fig_16dim}b. 
While we observe a strong diagonal contribution in the measurements, which represent the quality of correlations in the entangled state, the asymmetry can be attributed to the non-maximally entangled nature of the state in the measured pixel basis~\cite{Herrera2020} and alignment of the output modes to the collection single-mode fibers.

\textit{Discussion.---} 
The practical detection of high-dimensional entanglement presents a significant challenge in both theory and experiment. In this work, we have proposed a Schmidt number witness method based on a small number of measurements, each of which can be decomposed into a string of standard Pauli observables. This construction combines low measurement overhead with structural simplicity, while enabling Schmidt number detection both via fidelity estimation and beyond. We have experimentally demonstrated that these ideas lead to significant simplification in the circuit complexity required to realize the measurements. 
This complexity reduction is due to the fact that our high-dimensional measurements can be expressed as multiple parallel two-dimensional measurements. 
For $d=8$, our experiment relies on the implementation of four parallel $2$-dimensional transformations using MPLCs implemented with only two phase-planes. 
These measurements are easier than generic operations that typically require many more phase planes~\cite{goel_inverse_2024,kupianskyi_high-dimensional_2023,brandt_high-dimensional_2020,goel2023simultaneously,mothsara2026robust}.
Similarly, we implement our operations for $d=16$ as four parallel $4$-dimensional transformations with MPLCs using three planes. This is significantly fewer than a recent state-of-the-art experiment that used a $10$-plane MPLC for operations in the same dimension~\cite{lib2024resource}.

In general, we show that only $n$ unique pairings of optical modes are needed to realise a setup for $d=2^n$-dimensional entanglement certification.
The reduction in the required number of phase-layers~(and corresponding circuit depth) reduces the optical losses~\cite{fontaine2019laguerre}, alignment complexity~\cite{lib2025building}, and/or the number of optical components required to manufacture and stabilise the measurement devices. While we have demonstrated our technique with photonic states encoded in the transverse-spatial degree-of-freedom, an implementation with path-encoded photonic integrated circuits could be particularly interesting due to the scaling challenges associated with large-depth integrated circuits. In addition, the extension of our methods for entanglement certification to the multipartite regime and the detection of high-dimensionality in genuine multipartite entanglement is a natural next step \cite{Cobucci2024}.

\begin{acknowledgments}
We thank Shishir Khandelwal, Otfried G\"uhne and Nicola D'Alessandro  for discussions. We also thank Bohnishikha Ghosh for supporting with the experimental setup in the later stages of the experiment. This work is financially supported by the  Swedish Foundation for Strategic Research, by the Knut and Alice Wallenberg Foundation through the Wallenberg Center for Quantum Technology (WACQT), by the Swedish Research Council under Contract No.~2023-03498, by the Crafoord Foundation, the Krapperup Foundation and the Swedish Foundation for
Strategic Research. MM acknowledges financial support from the European Research Council (ERC) Starting and Consolidator Grants PIQUaNT (950402) and QUEST (101231627), the UK Engineering and Physical Sciences Research Council (EPSRC) (EP/Z533208/1, EP/Z533166/1), and the Royal Academy of Engineering Chair in Emerging Technologies programme (CiET-2223-112).

\end{acknowledgments}
\newpage
\bibliography{references_manuscript}

\appendix

\section{Anticommutation of Pauli strings}\label{AppA}
We show that all distinct pairs of Pauli strings anticommute. We prove this through induction. For $n=1$ the set of Pauli strings reduces to the three standard Pauli observables, which anticommute. Next, assume that pairwise anticommutation holds for $n-1$, i.e~ that
\begin{equation}\label{indassump}
	\{S_{n-1,i},S_{n-1,j}\}=0
\end{equation}
for any $i\neq j$. Now, consider the set of $n$-length strings $\{S_{n,j}\}_{j=1}^{2n+1}$. Select any  $i,j\in\{1,\ldots,2n-1\}$ such that  $i\neq j$. Then,
\begin{align}\nonumber
	\{S_{n,i},S_{n,j}\}&=(S_{n-1,i}\otimes Z)(S_{n-1,j}\otimes Z)\\
	&+(S_{n-1,j}\otimes Z)(S_{n-1,i}\otimes Z)\\
	&=\{S_{n-1,i},S_{n-1,j}\}\otimes Z=0,
\end{align}
where we used the induction assumption \eqref{indassump}. Next, choose $i\in\{1,\ldots,2n-1\}$, $j\in\{2n,2n+1\}$ and $A\in\{Y,Z\}$. Then,
\begin{align}
	\{S_{n,i},S_{n,j}\}&=(S_{n-1,i}\otimes Z)(\openone^{\otimes n}\otimes A)\\
	&+(\openone^{\otimes n}\otimes A)(S_{n-1,i}\otimes Z)\\
	&=S_{n-1,i}\otimes \{ZA,AZ\}=0,
\end{align}
where we used that distinct Pauli observables anticommute.

\section{Proof of Schmidt number bound}\label{AppB}
We prove the Schmidt number bounds for $W_n$ from the main text. To this end, let us define the set of states with Schmidt number at most $k$ by $\text{SN}(k)$  and the set of pure states with Schmidt rank at most $k$ by $\text{SR}(k)$. Also, we define the witness  operator  $B_n = \sum_{j=1}^{2n+1} S_{n,j} \otimes S_{n,j}^{T}$. Thus, we aim to place an upper bound on 
\begin{equation}
	\alpha(n,k) = 	\max_{\rho \in \text{SN}(k)} \tr(B_n \rho).
\end{equation}
Since the witness is linear, the value $\alpha(n,k)$ is achievable already with a pure state. Hence, we can restrict our analysis to pure states, $\ket{\psi}$ only. Thus,
\begin{equation}\label{upper_bound_sr_witness}
	\alpha(n,k) = 	\max_{\psi \in \text{SR}(k)} \tr(B_n \psi).
\end{equation}
Let us now obtain two different bounds for $\alpha(n,k)$, so the final upper bound is given by the minimum between these two.

\subsection{Fidelity-estimation bound}\label{AppB_2}
In this section we prove the bound \eqref{fidestimate}. Since we are aiming at inferring the Schmidt number via fidelity estimation, we relax the Schmidt number constraint to a fidelity constraint. Specifically, for any $\rho\in \text{SN}(k)$ it  holds that \cite{Terhal2000}
\begin{equation}\label{fidrel}
	\bracket{\phi^d}{\rho}{\phi^d}\leq \frac{k}{d}.
\end{equation} 
Let us denote the set of all states satisfying this constraint by $\mathcal{F}(k)$.

In the main text, we showed that the maximally entangled state $\ket{\phi^d}$, with $d=2^n$, is an eigenstate of $B_n$ with eigenvalue $2n+1$. Therefore, let us write the spectral decomposition of  $B_n$ as
\begin{equation}
	\begin{aligned}
		B_n &= (2n + 1) \phi^d + \sum_{i=2}^{4^n} \lambda_i \ketbra{\lambda_i}.
	\end{aligned}
\end{equation}
where $\lambda_i$ and $\ket{\lambda_i}$ are the remaining $4^n-1$ eigenvalues and eigenvectors.  Let us now write
\begin{equation}
	B_n = 4 \phi^d + \tilde{B}_n,
\end{equation}
where we have defined the shifted operator
\begin{equation}
\tilde{B}_n = (2n-3) \phi^d + \sum_{i=2}^{4^n} \lambda_i \ketbra{\lambda_i}.
\end{equation}
Assume momentarily that $\tilde{B}_n \preceq (2n-3)\openone$. Then, we have 
\begin{align}\nonumber
	\alpha(n,k)&\leq \max_{\rho \in \mathcal{F}(k)} \tr(B_n \rho)\\\nonumber
	&\leq \max_{\rho \in \mathcal{F}(k)} \Tr\left((4\phi^d+ (2n-3)\openone)\rho\right)\\\nonumber
	& =2n-3 +4\max_{\rho \in \mathcal{F}(k)}  \Tr\left(\phi^d \rho\right)\\
	& \leq 2n-3+\frac{4k}{d}=2n-3+\frac{k}{2^{n-2}},
\end{align}
where we used Eq~\eqref{fidrel}.

We now  need to prove that $\tilde{B}_n \preceq (2n-3)\openone$. This is achieved by induction. The statement is true for $n=1$, because
\begin{equation}
	\label{Phi+_Pauli-strings}
	\ketbra{\phi^+} = \frac{1}{4} \left[\openone \otimes \openone + X \otimes X - Y \otimes Y + Z \otimes Z\right]
\end{equation}
and therefore
\begin{equation}
	\tilde{B}_1 = \sum_{j=1}^{3} S_{1,j} \otimes S_{1,j}^{T} - 4\phi^+ = -\openone \otimes \openone.
\end{equation}
Now we only need to prove the validity of the bound for $n+1$ if we assume that it holds for $n$. We can rewrite $\tilde{B}_{n+1}$ as
\begin{equation}
	\begin{aligned}
		\tilde{B}_{n+1} &= \sum_{j = 1}^{2n + 3} S_{n+1,j} \otimes S_{n+1,j}^{T} - 4\, \phi^{d} \otimes \phi^+\\
		&= \sum_{j = 1}^{2n + 1} S_{n,j} \otimes Z \otimes S_{n,j}^{T} \otimes Z + \openone_n \otimes X \otimes \openone_n \otimes X\\
		&\quad - \openone_n \otimes Y \otimes \openone_n \otimes Y - 4\phi^d \otimes \phi^+.
	\end{aligned}
\end{equation}
By using \eqref{Phi+_Pauli-strings} and ordering the subspaces in all terms such that $(\mathcal{H}_{A_1} \otimes \dots \otimes \mathcal{H}_{A_n} \otimes \mathcal{H}_{B_1} \otimes \dots \otimes \mathcal{H}_{B_n}) \otimes \mathcal{H}_{A_{n+1}} \otimes \mathcal{H}_{B_{n+1}}$, we get
\begin{equation}
	\label{B_n+1}
	\begin{aligned}
		\tilde{B}_{n+1} = &\tilde{B}_n \otimes Z \otimes Z + \openone_n \otimes \openone_n \otimes X \otimes X\\
		& - \openone_n \otimes \openone_n \otimes Y \otimes Y- 4\phi^{d} \otimes \phi^+.
	\end{aligned}
\end{equation}
By the induction assumption we have  $\tilde{B}_n \preceq (2n-3)\mathds{1}$ which implies $ \tilde{B}_n \otimes Z\otimes Z \preceq (2n-3)\mathds{1}$. Since $\phi^{d} \otimes \phi^+$ is positive semidefinite we have 
\begin{align}\nonumber
	\tilde{B}_{n+1}&\preceq (2n-3)\mathds{1}+\openone_n\otimes \openone_n \otimes \left(X\otimes X-Y\otimes Y\right).
\end{align}
Since
\begin{equation}
	X\otimes X-Y\otimes Y \preceq 2 \openone_4
\end{equation}
it follows that 
\begin{equation}
\tilde{B}_{n+1} \preceq (2(n+1)-3) \openone.
\end{equation}

\subsection{Element-wise bound}\label{AppB_1}
In this section we prove the bound \eqref{alexbound}. First, notice that $\alpha(n,k)$ can be restated as the following maximization over matrices
\begin{equation}
\begin{aligned}
    \alpha(n,k)=&\max_{\psi\in\text{SR}(k)}\sum_{j=1}^{2n+1}\expval{S_{n,j}\otimes (S_{n,j})^\mathrm{T}}{\psi}\\
    =&\max_{\text{rank}(M)\leq
    k}\sum_{j=1}^{2n+1}\tr(MS_{n,j}M^\dagger S_{n,j}),
\end{aligned}
\end{equation}
where in the last step we have used Lemma 2.14 of \cite{Hjorungnes_2011}. In particular, $M$ is any matrix whose vectorization gives $\ket{\psi}$: $\text{vec}(M)=\ket{\psi}$. Moreover, since $\ket{\psi}$ can be any normalized state with Schmidt rank at most $k$, the maximization over $M$ refers to general complex matrices with $\|M\|_{HS}^2=1$ and rank at most $k$. We can further restrict the maximization only to positive semidefinite (PSD) $M$ matrices. To see this, let us consider the singular value decomposition (SVD) of $M$: $M=UD V^\dagger$. Then,
\begin{align}\label{eq:tr_M_1}
    &\sum_{j=1}^{2n+1}\tr(MS_{n,j}M^\dagger S_{n,j})=\\\nonumber
    &\sum_{j=1}^{2n+1}\tr((D^{1/2} V^\dagger S_{n,j}VD^{1/2})(D^{1/2} U^\dagger S_{n,j}UD^{1/2})).
\end{align}
Without loss of generality we can take
\begin{equation}
    \begin{aligned}
   &\sum_{j=1}^{2n+1}\tr(VD V^\dagger S_{n,i}VD V^\dagger S_{n,j})\leq\\
   &\sum_{j=1}^{2n+1}\tr(UD U^\dagger S_{n,j}UD U^\dagger S_{n,j}),
\end{aligned}
\end{equation}
so that by applying the Cauchy-Schwarz inequality in \eqref{eq:tr_M_1} and reordering the terms inside the traces, we get
\begin{equation}
\begin{aligned}
    &\sum_{j=1}^{2n+1}\tr(MS_{n,j}M^\dagger S_{n,j})\leq\\
    &\sum_{j=1}^{2n+1}\tr(UD U^\dagger S_{n,j}UD U^\dagger S_{n,j}).
\end{aligned}
\end{equation}
This implies that for any feasible $M=U D V^\dagger$ that is not PSD, we can find a feasible PSD matrix, $UD U^\dagger$, with at worst the same value of the objective function than $M$. Therefore,
\begin{equation}\label{eq:tr_M_2}
\alpha(n,k)=\max_{M\succeq0,\,\text{rank}(M)\leq k}\sum_{j=1}^{2n+1}\tr(MS_{n,j}M S_{n,j}),
\end{equation}
with $M=U D U^\dagger$. Next, let us consider the spectral decomposition $M=\sum_{a=1}^r d_a \ketbra{u_a}$. Here, $r\leq k$ is the rank of $M$, $\{d_a\}_a$ are its strictly positive eigenvalues, which satisfy $\sum_{a=1}^r d_a^2=1$ (because of $\|M\|_{HS}^2=1$) and $\{\ket{u_a}\}_a$ are the associated eigenvectors. Replacing this spectral decomposition, we have that the sum in \eqref{eq:tr_M_2} now reads
\begin{align}\nonumber\label{eq:tr_M_3}
    \sum_{j=1}^{2n+1}\tr(MS_{n,j}M S_{n,j})=&\sum_{a,b=1}^r d_a d_b \sum_{j=1}^{2n+1}|\mel{u_a}{S_{n,j}}{u_b}|^2\\
    =&\sum_{a,b=1}^r d_a d_b T_{ab},
\end{align}
where we have defined $T_{ab}=\sum_{j=1}^{2n+1}|\mel{u_a}{S_{n,j}}{u_b}|^2$. Assume for the moment that $T_{aa}\leq1$ and that $T_{ab}\leq2$ for $a\neq b$. Then, applying these bounds and the Cauchy-Schwarz inequality in \eqref{eq:tr_M_3} we get the desired upper bound:
\begin{equation}
\begin{aligned}
   &\sum_{a,b=1}^r d_a d_b T_{ab}\leq\sum_{a=1}^r d_a^2 +4\sum_{a<b} d_a d_b\\
   &=2\left(\sum_{a=1}^r d_a\right)^2 -\sum_{a=1}^r d_a^2\leq 2r-1\leq 2k-1,
\end{aligned}
\end{equation}
where we have used $\sum_{a=1}^r d_a^2=1$.

Let us check that actually $T_{aa}\leq1$ and $T_{ab}\leq2$ for $a\neq b$. For that purpose, let us first define the operator $S(\vec t)=\sum_{j=1}^{2n+1}t_j S_{n,j}$, where $\vec t$ is any $(2n+1)$-dimensional vector. Due to the anti-commutative relation of the $S_{n,j}$ matrices, the operator $S(\vec t)$ satisfies that
\begin{equation}
\begin{aligned}
        &S(\vec t)^2=\sum_{i,j} t_i t_j S_i S_j=\sum_j t_j^2\openone=\|\vec t\|_2^2\openone\\
    &\implies \|S(\vec t)\|_{\text{op}}=\|\vec t\|_2.
\end{aligned}
\end{equation}
For fixed $(a,b)$ let us define the vector $\vec w$ by $w_j=\mel{u_a}{S_{n,j}}{u_b}$. Then, $T_{ab}=\|\vec w\|_2^2$. Let us recall the variational definition of the 2-norm of a vector:
\begin{equation}\label{eq:norm_var}
\begin{aligned}
        \|\vec w\|_2=&\sup_{\|\vec t\|_2=1}\left|\sum_{j=1}^{2n+1} t_j^*w_j\right|\\
        =&\sup_{\|\vec t\|_2=1}\left|\mel{u_a}{\sum_{j=1}^{2n+1}t_j^* S_{n,j}}{u_b}\right|.
\end{aligned}
\end{equation}
Now, let us distinguish the aforementioned cases over $(a,b)$. If $a=b$, the vector $\vec w$ is a real vector and $\vec t$ in \eqref{eq:norm_var} can also be chosen to be real. Hence, 
\begin{equation}
\begin{aligned}
        \|\vec w\|_2=&\sup_{\|\vec t\|_2=1}\left|\mel{u_a}{\sum_{j=1}^{2n+1}t_j S_{n,j}}{u_a}\right|\\
        =&\sup_{\|\vec t\|_2=1}\left|\mel{u_a}{S(\vec t)}{u_a}\right|\\
        \leq&\sup_{\|\vec t\|_2=1}\|S(\vec t)\|_{\text{op}}=\sup_{\|\vec t\|_2=1}\|\vec t\|_2=1.
\end{aligned}
\end{equation}
Thus, $T_{aa}= \|\vec w\|_2^2\leq 1$. If, on the contrary, $a\neq b$, we take $\vec t=\vec x - i\vec y$
and the condition $\|\vec t\|_2=1$ is equivalent to $\|\vec x\|_2^2+\|\vec y\|_2^2=1$. In consequence,
\begin{align}\nonumber
    &\left|\sum_{j=1}^{2n+1} t_j^*w_j\right|=\left|\mel{u_a}{\sum_{j=1}^{2n+1}x_j S_{n,j}}{u_b}+i\mel{u_a}{\sum_{j=1}^{2n+1}y_j S_{n,j}}{u_b} \right|\\\nonumber
    &\leq\sqrt{2}\sqrt{\left|\mel{u_a}{\sum_{j=1}^{2n+1}x_j S_{n,j}}{u_b}\right|^2+\left|\mel{u_a}{\sum_{j=1}^{2n+1}y_j S_{n,j}}{u_b} \right|^2}\\
    &=\sqrt{2}\sqrt{\|S(\vec x)\|_{\text{op}}^2+\|S(\vec y)\|_{\text{op}}^2}\\\nonumber
    &=\sqrt{2}\sqrt{\|\vec x\|_2^2+\|\vec y\|_2^2}=\sqrt{2}.
\end{align}
Thus, $T_{ab}= \|\vec w\|_2^2\leq 2$ for $a\neq b$.

\section{Numerical search}\label{AppC}

In this Appendix, we discuss the numerical method used to search for the value $\alpha(n,k)$ defined in \eqref{upper_bound_sr_witness}. By enlarging the Hilbert space $\mathcal{H}_{A} \otimes \mathcal{H}_{B}$ to $(\mathcal{H}_{A} \otimes \mathcal{H}_{B}) \otimes (\mathcal{H}_{A'}^{k} \otimes \mathcal{H}_{B'}^{k})$, where $\dim(\mathcal{H}_{A'}^{k}) = \dim(\mathcal{H}_{B'}^{k}) = k$, the values of $\alpha(n,k)$ for a given Schmidt rank $k$ can be found addressing a separability problem \cite{Hulpke2004}, i.e.
\begin{equation}\label{witness_enlarged_expression}
	\alpha(n,k) = \max_{\sigma \in \text{SEP}(AA'|BB')} k^2 \, \Tr((B_{n})_{AB} \otimes \ketbra{\phi_{k}^{+}}_{A'B'} \sigma),
\end{equation}
where $\ket{\phi_{k}^{+}}=\frac{1}{\sqrt{k}}\sum_{p=1}^{k}\ket{pp}$ and $\sigma_{AA'BB'}= \varphi_{AA'} \otimes \chi_{BB'}$ is a separable state across the bipartition $AA'|BB'$ with $\Tr_{B}(\chi_{BB'}) = \frac{1}{k}\openone$. To show this result, express a generic state $\ket{\psi}\in \text{SR}(k)$ in its Schmidt decomposition, $\ket{\psi} = \sum_{i=1}^{k}\lambda_{i}\ket{a_{i} b_{i}}$, where $\lbrace \lambda_{i} \rbrace_{i}$ are the Schmidt coefficients such that $\sum_{i=1}^{k}\lambda_{i}^{2} = 1$ and $\lbrace \ket{a_i},\ket{b_i} \rbrace_i$ are two orthonormal sets of vectors on $\mathcal{H}_{A}$ and $\mathcal{H}_{B}$, respectively. Therefore,
\begin{widetext}
\begin{equation}\label{witness_expanded_proof}
	\begin{aligned}
	\Tr(B_{n}\psi) &= \sum_{p,q=1}^{k} \lambda_{p}^{*}\lambda_{q} \bra{a_{p} b_{p}}B_{n}\ket{a_{q}b_{q}} =  \sum_{i,j,l,m,p,q=1}^{k} \lambda_{i}^{*}\lambda_{l} \bra{a_{i} b_{j}}B_{n}\ket{a_{l}b_{m}} \underbrace{\delta_{i,p}\delta_{j,p}}_{=\braket{ij}{pp}} \underbrace{\delta_{l,q}\delta_{m,q}}_{=\braket{qq}{lm}}\\
	&= k^2 \sum_{i,j,l,m=1}^{k} \frac{1}{k}\lambda_{i}^{*}\lambda_{l} \bra{a_i b_{j}}_{AB} \otimes \bra{i j}_{A'B'} \left[ (B_{n})_{AB} \otimes \frac{1}{k}\sum_{p,q=1}^{k}\ketbra{pp}{qq}_{A'B'}\right] \ket{a_{l}b_{m}}_{AB} \otimes \ket{lm}_{A'B'}\\
	&= k^2 \left(\sum_{i=1}^{k} \lambda_{i}^{*} \bra{a_i i}_{AA'} \otimes \frac{1}{\sqrt{k}}\sum_{j=1}^{k} \bra{b_j j}_{BB'} \right)\left[ (B_{n})_{AB} \otimes \frac{1}{k}\sum_{p,q=1}^{k}\ketbra{pp}{qq}_{A'B'}\right] \left(\sum_{l=1}^{k} \lambda_{l} \ket{a_l l}_{AA'} \otimes \frac{1}{\sqrt{k}}\sum_{m=1}^{k} \ket{b_m m}_{BB'}\right),
	\end{aligned}
\end{equation}
\end{widetext}
where we can define $\ket{\varphi}_{AA'} = \sum_{i=1}^{k} \lambda_i \ket{a_i i}_{AA'}$ and $\chi_{BB'} = \frac{1}{\sqrt{k}}\sum_{j=1}^{k}\ket{b_j j}_{BB'}$ to recover the result in \eqref{witness_enlarged_expression}.

Note that by definition $\varphi_{AA'}$ and $\chi_{BB'}$ are states, i.e.~they satisfy $\varphi, \chi \succeq 0$ and $\Tr(\varphi) = \Tr(\chi) = 1$. Moreover, from \eqref{witness_expanded_proof} we also require that $\Tr_{B}(\chi) = \frac{1}{k} \openone$. Therefore, we can write the optimization problem as follows 
\begin{equation}
	\begin{aligned}
	\max_{\lbrace \varphi, \chi \rbrace} & \quad k^2 \, \Tr((B_{n})_{AB} \otimes \ketbra{\phi_{k}^{+}}_{A'B'} \varphi_{AA'} \otimes \chi_{BB'} )\\
	\text{s.t.} &\quad \varphi \succeq 0, \quad \Tr(\varphi) = 1,\\
	&\quad \chi \succeq 0, \quad \Tr(\chi) = 1, \quad \Tr_{B}(\chi) = \frac{1}{k}\openone.
	\end{aligned}
\end{equation}
This problem can be relaxed using a numerical search algorithm with the following steps:
\begin{enumerate}
	\item  Select a random state $\chi_{BB'}$. Then, the following optimisation
	\begin{equation}\label{optimization_phi}
		\begin{aligned}
			\max_{\lbrace \varphi\rbrace} & \quad k^2 \, \Tr((B_{n})_{AB} \otimes \ketbra{\phi_{k}^{+}}_{A'B'} \varphi_{AA'} \otimes \chi_{BB'} )\\
			\text{s.t.} &\quad \varphi_{AA'} \succeq 0, \quad \Tr(\varphi_{AA'}) = 1.
		\end{aligned}
	\end{equation}
    can be performed by selecting $\varphi_{AA'}$ to be the eigenvector which corresponds to the largest eigenvalue of the operator $k^2 \,\tr_{BB'}((B_{n})_{AB} \otimes \ketbra{\phi_{k}^{+}}_{A'B'})(\openone_{AA'}\otimes \chi_{BB'}))$.
	\item Use the optimal state $\varphi_{AA'}$ obtained in the previous step and optimise over $\chi_{BB'}$ using semidefinite programming relaxation (SDP) \cite{Tavakoli2024}
	\begin{equation}
		\begin{aligned}
			\max_{\lbrace \chi\rbrace} & \quad k^2 \, \Tr((B_{n})_{AB} \otimes \ketbra{\phi_{k}^{+}}_{A'B'} \varphi_{AA'} \otimes \chi_{BB'} )\\
			\text{s.t.} &\quad \chi_{BB'} \succeq 0, \quad \Tr(\chi_{BB'}) = 1, \quad \Tr_{B}(\chi_{BB'}) = \frac{1}{k}\openone.
		\end{aligned}
	\end{equation}
	\item Return to step 1 and use the optimal $\chi_{BB'}$ from the previous step as an input for the optimization \eqref{optimization_phi}.
\end{enumerate}

The procedure can be repeated until desired convergence is reached. Our implementation can be found at \cite{github-code}.

\section{Subfamily maximization and eigenvalue-of-subgraph relation}\label{AppD}

Let us proof the maximization solution in \eqref{Subset_Opt}. To this end, we again use the result proven in Appendix \ref{AppB_1}:
\begin{equation}
\alpha (n,k)=\max_{M\succeq0,\,\text{rank}(M)\leq k}\sum_{j=1}^{2n+1}\tr(MS_{n,j}M S_{n,j}).
\end{equation}
We take the set of PSD diagonal matrices, $M\equiv D \succeq0$, as the subfamily to maximize over:
\begin{equation}\label{eq:sum1}
\max_{D\succeq0,\,\text{rank}(D)\leq k}\sum_{j=1}^{2n+1}\tr(D S_{n,j} D S_{n,j}).
\end{equation}
The set of matrices we are maximizing over corresponds, via vectorization, to the set $T_k$ introduced in the main text. Let us focus on the $j=2n+1$ term, which we take to be $S_{n,2n+1}=Z^{\otimes n}$. In this case, both $D$ and $S_{n,2n+1}$ are diagonal and commute. Then, $\tr(D S_{n,2n+1} D S_{n,2n+1})=\tr(D^2)=1$, where we used $ S_{n,2n+1}^2 =\openone$. The sum in Eq~\eqref{eq:sum1} now reads
\begin{equation}\label{eq:sum2}
    1+\sum_{j=1}^{2n}\tr(D S_{n,j} D S_{n,j}).
\end{equation}
The other generators, $S_{n,j}$ for $j=1,\dots,2n$, can be relabelled so that
\begin{equation}
    \{S_{n,j}\}_{j=1}^{2n}=\{X_l,Y_l\}_{l=1}^n,
\end{equation}
where 
\begin{equation}
\begin{aligned}
    &X_l=\openone^{\otimes (l-1)}\otimes X\otimes Z^{\otimes (n-l)},\\
    &Y_l=\openone^{\otimes (l-1)}\otimes Y\otimes Z^{\otimes (n-l)}.
\end{aligned}
\end{equation}
Then, the operators $\{X_l,Y_l\}$ act non-diagonally only over the $l$ qubit. In particular, for a computational basis element of the form $\ket{\boldsymbol{a}}=\ket{a_1 a_2\cdots a_n}\in\{0,1\}^n$, we have $X_l \ketbra{\boldsymbol{a}} X_l^\dagger=Y_l \ketbra{\boldsymbol{a}} Y_l^\dagger=\ketbra{\boldsymbol{a}\oplus \boldsymbol{e}_l}$, with $\ket{\boldsymbol{e}_l}=\ket{0\cdots 0 1 0\cdots 0}$ the basis element whose only excited component corresponds to the $l$ qubit. 

Let us express the diagonal matrix in the computational basis: $D=\sum_{\boldsymbol{a}} d_{\boldsymbol{a}}\ketbra{\boldsymbol{a}}$. Then $X_lD X_l^\dagger=Y_lD Y_l^\dagger=\sum_{\boldsymbol{a}} d_{\boldsymbol{a}}\ketbra{\boldsymbol{a}\oplus \boldsymbol{e}_l}$ and 
\begin{equation}
\begin{aligned}
    &\tr(D X_l D X_l)= \tr(D Y_l D Y_l)\\
    &=\sum_{\boldsymbol{a},\boldsymbol{b}} d_{\boldsymbol{a}}d_{\boldsymbol{b}}\delta_{\boldsymbol{a}\oplus \boldsymbol{e}_l,\, \boldsymbol{b}}=\sum_{\boldsymbol{a}} d_{\boldsymbol{a}}d_{\boldsymbol{a}\oplus \boldsymbol{e}_l}.
\end{aligned}
\end{equation}
Thus, the sum in Eq.~\eqref{eq:sum2} becomes
\begin{equation}
    1+\sum_{j=1}^{2n}\tr(D S_{n,j} D S_{n,j})=1+2 \sum_{\boldsymbol{a}}\sum_{l=1}^n d_{\boldsymbol{a}}d_{\boldsymbol{a}\oplus \boldsymbol{e}_l}.
\end{equation}
This last term coincides with $\vec d^{\ \rm{T}} \text{Ad}(Q_n) \vec d$, where the matrix $\text{Ad}(Q_n)$ refers to the adjacency matrix of the $n$-qubit hypercube, i.e., the matrix satisfying $\mel{\boldsymbol{a}}{\text{Ad}(Q_n)}{\boldsymbol{b}}=1$ if the qubit string $\ket{\boldsymbol{b}}=\ket{\boldsymbol{a}\oplus \boldsymbol{e}_l}$ for some $l$ and $\mel{\boldsymbol{a}}{\text{Ad}(Q_n)}{\boldsymbol{b}}=0$ in any other case. Finally,
\begin{align}
     &\max_{D\succeq 0,\,\text{rank}(D)\leq k}\sum_{j=1}^{2n+1}\tr(D S_{n,j} D S_{n,j})\\\nonumber
     &=1+2 \max_{\vec d}\vec d^{\ \rm{T}} \text{Ad}(Q_n) \vec d= 1+2 \max_{|S|= k} \lambda_{\max}(\text{Adj}(Q_n[S])),
\end{align}
where in the last step we have maximized over all vectors $\vec d$ verifying $\sum_{\boldsymbol{a}} d_{ \boldsymbol{a} }^2=1$ and with at most $k$ non-zero entries. 

Let us denote $\mu(n,k)=\max_{|S|= k} \lambda_{\max}(\text{Adj}(Q_n[S]))$. In Table~\ref{Tab2} we show the value $1+2\mu(n,k)$ for different choices of $n,k$. In particular we take $n\leq5$ and let $k=1,\dots,n$. From Table~\ref{Tab2}, one may guess that if $k=2^m$ for $m=0,\dots, n$ then $\mu(n,2^m)= m$. Nevertheless, this is just accidental for low values of $n$. The actual relation reads $\mu(n,2^m)\geq m$, and the lower bound is always accomplished by taking $S=Q_m$. Even though for small $n$ the inequality relation turns out to be an equality, this is not always the case. For instance, if $n=24$ and $m=5$, we can take as the subgraph $S$ the one generated by
\begin{equation}
    S=\{\emptyset\} \cup \{\{i\}: 1\leq i\leq 24\} \cup \{\{1,i\}: 2\leq i\leq 8\}.
\end{equation} 
This subgraph fulfils that $|S|=32=2^5$. Now notice that a partition, $C$, of the subgraph $S$ into cells is 
\begin{equation}
\begin{aligned}
    &C_0=\{\emptyset\},\quad C_1=\{\{1\}\},\\
    &C_2=\{\{i\}: 2\leq i\leq 8\},\quad C_3=\{\{i\}: 9\leq i\leq 24\},\\
    &C_4=\{\{1,i\}: 2\leq i\leq 8\}.
\end{aligned}
\end{equation}
This is an equitable partition, i.e. for any two cells $C_k, C_l$ all the vertices in $C_k$ must be connected with the same number of vertices in $C_l$. By Lemma 2.3.1 of \cite{BrouwerHaemers2012}, we know that the eigenvalues of the quotient matrix $Q_C$ of the equitable partition $\{C_k\}_k$ are also eigenvalues of $\text{Adj}(Q_n[S])$. An entry in the quotient matrix, $(Q_C)_{k,l}$, gives the number of connections between the elements in $C_k$ and $C_l$ divided by the number of elements of $C_k$ for $k,l=0,\dots,4$. In particular, 
\begin{equation}
   Q_C= \left(
    \begin{array}{ccccc}
        0 & 1 & 7 & 16 & 0 \\
        1 & 0 & 0 & 0 & 7 \\
        1 & 0 & 0 & 0 & 1 \\
        1 & 0 & 0 & 0 & 0 \\
        0 & 1 & 1 & 0 & 0 
    \end{array}
    \right),
\end{equation}
and its largest eigenvalue is $\sqrt{16 +2 \sqrt{23}}$. Therefore, $\mu(n,k)\geq \lambda_{\max}\text{Adj}(Q_n[S])\geq \sqrt{16 +2 \sqrt{23}}>5$.

\setlength{\tabcolsep}{3pt}
\renewcommand{\arraystretch}{1.15}
\begin{table*}[t]
	\centering
	\small

	\begin{tabular}{c||c|c|c|c@{\qquad}|c||c}
		$k$
		& $n=2$
		& $n=3$
		& $n=4$
		& $n=5$
		& $k$
		& $n=5$ \\ \hline\hline

		1  & $1^*$ & $1^*$ & $1^*$ & $1^*$
		& 17 & $9.0333$ \\ \hline

		2  & $3^*$ & $3^*$ & $3^*$ & $3^*$
		& 18 & $1+\sqrt{38+2\sqrt{57}}$ \\ \hline

		3  & $1+2\sqrt{2}$ & $1+2\sqrt{2}$ & $1+2\sqrt{2}$ & $1+2\sqrt{2}$
		& 19 & $9.1557$ \\ \hline

		4  & $5^*$ & $5^*$ & $5$ & $5$
		& 20 & $5+\sqrt{10+2\sqrt{17}}$ \\ \hline

		5  & $-$ & $1+\sqrt{10+2\sqrt{17}}$ & $1+\sqrt{10+2\sqrt{17}}$ & $1+\sqrt{10+2\sqrt{17}}$
		& 21 & $9.3436$ \\ \hline

		6  & $-$ & $3+2\sqrt{2}$ & $3+2\sqrt{2}$ & $3+2\sqrt{2}$
		& 22 & $9.4686$ \\ \hline

		7  & $-$ & $1+2\sqrt{7}$ & $1+2\sqrt{7}$ & $1+2\sqrt{7}$
		& 23 & $9.6026$ \\ \hline

		8  & $-$ & $7^*$ & $7^*$ & $7$
		& 24 & $7+2\sqrt{2}$ \\ \hline

		9  & $-$ & $-$ & $1+\sqrt{22+2\sqrt{57}}$ & $1+\sqrt{22+2\sqrt{57}}$
		& 25 & $9.8916$ \\ \hline

		10 & $-$ & $-$ & $3+\sqrt{10+2\sqrt{17}}$ & $3+\sqrt{10+2\sqrt{17}}$
		& 26 & $9.9966$ \\ \hline

		11 & $-$ & $-$ & $7.4686$ & $7.4686$
		& 27 & $10.1083$ \\ \hline

		12 & $-$ & $-$ & $5+2\sqrt{2}$ & $5+2\sqrt{2}$
		& 28 & $5+2\sqrt{7}$ \\ \hline

		13 & $-$ & $-$ & $7.9966$ & $7.9966$
		& 29 & $10.3983$ \\ \hline

		14 & $-$ & $-$ & $3+2\sqrt{7}$ & $3+2\sqrt{7}$
		& 30 & $3+2\sqrt{8+2\sqrt{10}}$ \\ \hline

		15 & $-$ & $-$ & $1+2\sqrt{8+2\sqrt{10}}$ 
		   & $1+2\sqrt{8+2\sqrt{10}}$
		& 31 & $1+2\sqrt{15+2\sqrt{19}}$ \\ \hline

		16 & $-$ & $-$ & $9^*$ & $9^*$
		& 32 & $11^*$ 
	\end{tabular}

	\caption{Values of $1+2\mu(n,k)$ for $n\leq 5$. We mark with an asterisk those values matching the upper bounds from Eq.~\eqref{fidestimate} and Eq.~\eqref{alexbound}.}
	\label{Tab2}
\end{table*}

\section{Block-diagonalisability of Pauli strings\label{App_block_diagonal}}
We prove that for a fixed dimension $d=2^n$, the set of Pauli strings $\{S_{n,j}\}_j$ and their eigenvectors can always be recast as a permutation acting over a block-diagonal matrix. Moreover, the total number of permutations used for block-diagonalising the whole set is $n$. The main idea behind the proof relies on noticing that we can relabel the Pauli strings so that
\begin{equation}
    \{S_{n,j}\}_{j=1}^{2n+1}=\{X_l,Y_l\}_{l=1}^n\cup \{Z^{\otimes n}\},
\end{equation}
where 
\begin{equation}
\begin{aligned}
    &X_l=\openone^{\otimes (l-1)}\otimes X\otimes Z^{\otimes (n-l)},\\
    &Y_l=\openone^{\otimes (l-1)}\otimes Y\otimes Z^{\otimes (n-l)}.
\end{aligned}
\end{equation}
Since $Z^{\otimes n}$ is already block-diagonal (actually, diagonal), we only need to deal with $X_l,Y_l$. Hence, we require to interfere every pair of modes whose binary representation differs in exactly one place. These come in $n$ groups of pairings required to realise each single $X_l$ ($Y_l$) at position $l$, where the $2^{n-1}$ pairs correspond to the remaining $k \neq l$ substrings.

A detailed mathematical proof of the result is as follows: let us define the set of permutations $\{R_l\}_{l=1}^n$ as the permutations taking the $l$ qubit and placing it in the last position of the qubit-string, i.e.
\begin{equation}
    R_l \ket{x_1 x_2\cdots x_{l-1} x_l x_{l+1}\cdots x_n}=\ket{x_1 x_2\cdots x_{l-1} x_{l+1}\cdots x_n x_l} 
\end{equation}
Then, we have
\begin{equation}
\begin{aligned}
    R_l X_l R_l^\dagger &= R_l(\openone^{\otimes (l-1)}\otimes X\otimes Z^{\otimes (n-l)})R_l^\dagger\\
    &=\openone^{\otimes (l-1)}\otimes Z^{\otimes (n-l)}\otimes X,\\
    R_l Y_l R_l^\dagger &= R_l(\openone^{\otimes (l-1)}\otimes Y\otimes Z^{\otimes (n-l)})R_l^\dagger\\
    &=\openone^{\otimes (l-1)}\otimes Z^{\otimes (n-l)}\otimes Y.
\end{aligned}
\end{equation}
In both cases, the last expression is indeed block-diagonal with blocks of size 2, so 
\begin{equation}
\begin{aligned}
    X_l&= R_l^\dagger (\openone^{\otimes (l-1)}\otimes Z^{\otimes (n-l)}\otimes X) R_l,\\
    Y_l&= R_l^\dagger (\openone^{\otimes (l-1)}\otimes Z^{\otimes (n-l)}\otimes Y) R_l,
\end{aligned}
\end{equation}
and we get the desired decomposition for the $X_l,Y_l$ matrices. 
Let us now call $U_X$ and $U_Y$ the unitary matrices diagonalizing $X$ and $Y$ respectively: $U_X^\dagger X U_X =Z$ and $U_Y^\dagger Y U_Y =Z$. We define the unitary matrices 
\begin{equation}
    \begin{aligned}
        U_{X,l}&=R_l (\openone^{\otimes (n-1)}\otimes U_X)R_l^\dagger,\\
        U_{Y,l}&=R_l (\openone^{\otimes (n-1)}\otimes U_Y)R_l^\dagger.
    \end{aligned}
\end{equation}
These matrices are permutations of block-diagonal matrices and it is direct to check that they diagonalise the set of Pauli strings:
\begin{equation}
\begin{aligned}
    U_{X,l}^\dagger X_l U_{X,l}&=(\openone^{\otimes (n-1)}\otimes U_X^\dagger) (R_l X_l R_l^\dagger)(\openone^{\otimes (n-1)}\otimes U_X)\\
    &=\openone^{\otimes (l-1)}\otimes Z^{\otimes(n-l)}\otimes U_X^\dagger X U_X\\
    &=\openone^{\otimes (l-1)}\otimes Z^{\otimes(n-l+1)},\\
    U_{Y,l}^\dagger Y_l U_{X,l}&=(\openone^{\otimes (n-1)}\otimes U_X^\dagger) (R_l Y_l R_l^\dagger)(\openone^{\otimes (n-1)}\otimes U_Y)\\
    &=\openone^{\otimes (l-1)}\otimes Z^{\otimes(n-l)}\otimes U_Y^\dagger Y U_Y\\
    &=\openone^{\otimes (l-1)}\otimes Z^{\otimes(n-l+1)}.
\end{aligned}
\end{equation}
Therefore, the matrices $U_{X,l},U_{Y,l}$ (which are of the desired form) have as their columns the eigenvectors of the corresponding $X_l,Y_l$ Pauli strings. Finally, since $Z^{\otimes n}$ is diagonal, their eigenvectors can be taken to be the computational basis and the associated unitary matrix is $U_{Z^{\otimes n}}=\openone$, which is already diagonal.

\begin{figure*}[t]
 	\centering
 	\includegraphics[width=1.3\columnwidth]{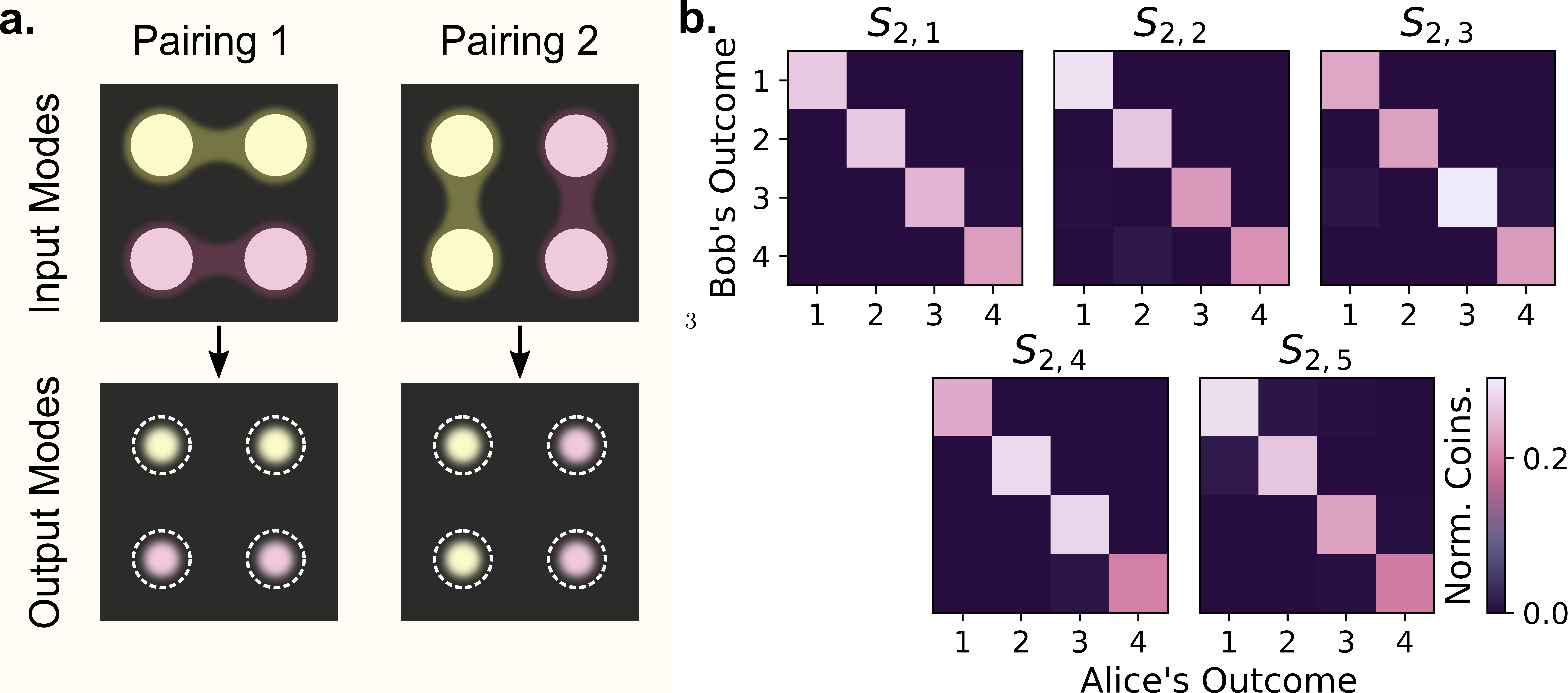}
 	\caption{(a) The four input pixels~(and output spots) are arranged in a grid configuration that allows for nearest neighbour operations in two different permutations ($R_j$), effectively needing two parallel two-mode operations. (b) Measured normalised coincidence (Norm. Coins.) matrices between Alice's and Bob's outcomes corresponding to each Pauli-string observable for $d=4$ dimensional state. }\label{fig_4dim}
\end{figure*}

\section{Experimental Details\label{App_Exp}}

The pair of entangled photons in the telecom band are generated using the process of type-II spontaneous parametric downconversion~(SPDC) pumped with a $775 nm$ continuous wave~(CW) pump laser with a power of $400 mW$.
The Gaussian-shaped pump beam is appropriately shaped using a telescope, allowing for high-dimensional spatial entanglement to be generated~\cite{srivastav2022characterizing,Herrera2020}. 
The remaining pump after the downconversion is filtered using a dichroic mirror as well as a low-pass filter.
The filtered emission, i.e. the bi-photons, propagate through a lens, and are then spatially separated using a polarising beam-displacer and incident on a multi-plane light converter~(MPLC), such that the first plane of reflection is placed at the back-focal plane. 
The MPLC is implemented using a spatial-light modulator~(SLM, Holoeye PLUTO-2.1-TELECO-142) and a mirror parallel to each other.
The MPLC is aligned such that both the spatially separated (vertically displaced) photons reflect on the SLM thrice with a free-space propagation of $5.4 cm$ in between. To implement a two-plane MPLC, we simply turn-off the area on the SLM where the second reflection takes place, effectively making the propagation distance between the two planes to be $10.8 cm$.
We program a given unitary operation within the MPLCs using an inverse-design algorithm called wavefront-matching~\cite{hashimoto_optical_2005,sakamaki_new_2007}. The wavefront-matching algorithm is used to iteratively couple each input mode transformed according the implemented unitary, to an output mode, by the means of updating the intermediate phase-planes.
After being spatially transformed within the MPLCs, the two photons are aligned to be coupled into two single-mode cores of a multi-core fiber, which are placed on the back-focal plane of a coupling lens. 
While in principle, the implemented transformations can be used for performing multi-outcome measurements, one requires an array of single-photon detectors to achieve the same. 
Here, we assume fair-sampling and measure the joint-click statistics of each outcome for Alice and Bob sequentially by redirecting the outcomes to be aligned to the aforementioned single-mode cores using the final plane of each MPLC. 
Each core of the fiber is connected to an superconducting-nanowire single-photon detectors~(SNSPDs) for photo detection.
The detection events are correlated using a time-tagger unit~(Swabian TimeTagger Ultra) with a correlation window of $0.6~ns$.

\begin{figure*}[b!]
 	\centering
 	\includegraphics[width=1.9\columnwidth]{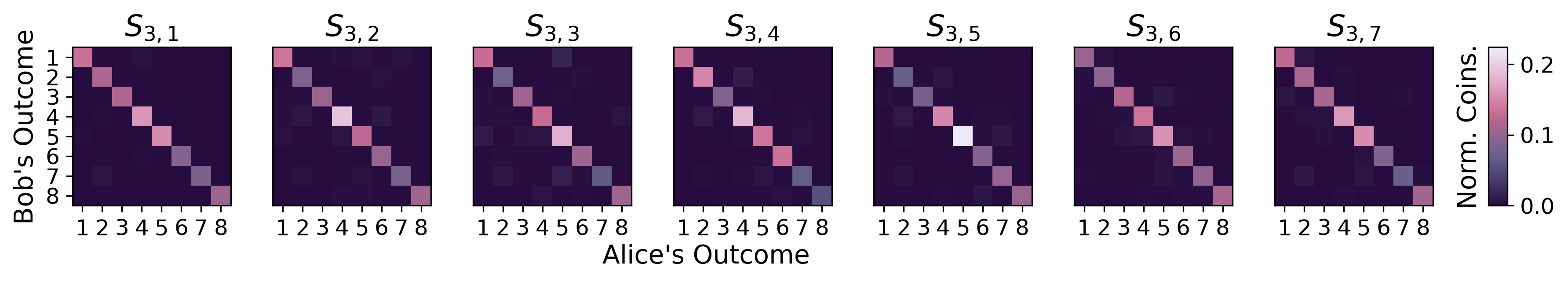}
 	\caption{Measured normalised coincidence (Norm. Coins.) matrices between Alice's and Bob's outcomes corresponding to each Pauli-string observable for $d=8$ dimensional state.
    }\label{fig_8dim_res}
 \end{figure*}

\section{Additional Data\label{App_data}}
Measured data, as well as mode arrangements chosen for $(n,d)=(2,4)$, are shown in Fig.~\ref{fig_4dim}. For $n=2$, only two unique permutations allow for local two-mode operations over $5$ Pauli-string observations. Geometrically, we choose a simple grid of 4 modes that allow for these operations as shown in Fig.~\ref{fig_4dim}a.  
Measured data for the case of $(n,d)=(3,8)$ is also shown in Fig.~\ref{fig_8dim_res}.

\end{document}